\documentclass[conference]{IEEEtran}

\IEEEoverridecommandlockouts

\usepackage{cite}
\usepackage{amsmath,amssymb,amsfonts}
\usepackage{graphicx}
\usepackage{booktabs}
\usepackage{multirow}
\usepackage{xcolor}
\usepackage{url}
\usepackage{hyperref}
\usepackage{algorithm}
\usepackage{algpseudocode}
\usepackage{booktabs}
\usepackage{tabularx}

\begin{document}

\title{Guarded Equivalence Predicates for Scalable Formal Hardware Information-Flow Verification
}

\author{%
\IEEEauthorblockN{Anonymous Submission}
}

\author{
\IEEEauthorblockN{Liangtao Dai, Yimin Gao, Melika Morsali and Mircea R. Stan}
\IEEEauthorblockA{
Department of Electrical and Computer Engineering\\
University of Virginia\\
Charlottesville, VA, USA\\
\{kvf4sf, yg9bq, qfc2zn, mircea\}@virginia.edu
}
}

\maketitle

\begin{abstract}

Formal hardware information-flow verification is a principled way to rule out secret-dependent functional or timing observations, but scaling such proofs remains difficult. Self-composition reduces information-flow verification to safety checking over two circuit copies, creating relational proof obligations that are hard for a generic PDR engine to discover from bit-level logic alone. Recent PDR-based techniques exploit this duplicated structure through copy symmetry and global cross-copy equivalence predicates. These predicates are effective when corresponding internal signals agree throughout the reachable state space, but they do not capture equalities that are relevant only in a specific control context. We observe that such contextual relations arise naturally in hardware IFV proofs: an internal signal pair may need to agree only within a control phase, transaction window, loop state, or protocol region. We introduce guarded equivalence predicates to expose these relations to PDR. Rather than treating a proposed contextual equality as an assumption, the verifier submits the corresponding mismatch condition as an auxiliary blocking obligation. Guards are proposed from relational counterexamples-to-induction using CTI-local extraction and state-split search; only candidates proved unreachable by the backend affect the proof. Across 12 IFV benchmarks and two PDR backends, guarded predicates convert two contextual baseline timeouts into completed proofs within 34.2--89.5s under an 1800s limit, while reducing proof time by up to 10.8$\times$ on additional benchmarks.

\end{abstract}


\section{Introduction}

Hardware information-flow verification (IFV) proves that protected sources do not influence attacker-observable sinks, including functional outputs and timing-observable behavior~\cite{Gleissenthall2021constranhwverification}.
Such guarantees are important because hardware enforces confidentiality, privilege boundaries, and trusted execution, while hardware vulnerabilities can remain exploitable even when software is correctly written ~\cite{Lou2022hw_security_attacks,Kocher2019SepctreMeltdown}. Formal IFV addresses this problem by proving security over all executions of a design, rather than relying only on simulation, testing, fuzzing, or manually selected traces~\cite{Hu2022HIFT,Ardeshiricham2017RTLIFT,Hu2011GLIFT,Gleissenthall2019IODINE}.

A standard way to verify non-interference is self-composition ~\cite{Barthe2004selfcomposition}. The verifier builds two copies of the same hardware design, constrains their public inputs to match, allows their secret inputs to differ, and checks that corresponding observable outputs remain equal. This reduces a hyperproperty over pairs of executions to a safety property over a single composed transition system ~\cite{Clarkson2010Hyperproperty}. Although this reduction is general and direct, it also creates a relational proof obligation over two structurally identical circuit copies. Generic model checkers may miss this structure.

Recent work has shown that PDR-based hardware IFV can benefit substantially from the duplicated structure created by self-composition. In particular, SecIC3~\cite{tan2026secic3} customizes IC3/PDR with copy-swap symmetry and cross-copy equivalence predicates, giving the proof engine a relational vocabulary for reasoning about corresponding signals in the two circuit copies. This approach is effective on many self-composed IFV instances, but it can still struggle on designs whose security proof is entangled with complex control behavior, such as multi-cycle arithmetic units and FSM-rich protocol regions.

We observe that, in such cases, the missing proof fact is often not simply a global cross-copy equality. Instead, the useful relation is tied to a control context: two corresponding internal signals may need to agree only within a particular FSM state, loop state, transaction window, or protocol phase. Outside that context, the equality may be irrelevant to the observable behavior, false on reachable states, or too expensive for PDR to establish as an unconditional invariant. This suggests that the verifier should not only have access to cross-copy equivalence predicates, but should also be able to test whether a mismatch is unreachable under a specific control context.

We introduce \emph{guarded equivalence predicates} for PDR-based hardware IFV. Given a candidate control context (g) and a cross-copy mismatch predicate ($\mathit{neq}_X$), our method submits the guarded mismatch cube ($g \wedge \mathit{neq}_X$) to the ordinary PDR blocking procedure. If PDR proves this region unreachable, the learned clauses become part of the proof and establish the corresponding guarded equality on reachable states. If the region is reachable, or if the backend cannot block it, the candidate is discarded. Thus guarded predicates are not assumptions; they are auxiliary proof obligations accepted only by the backend's proof-producing mechanism.


The remaining question is how to find useful guards automatically. Our method uses information already produced inside PDR proof search. CTI-local search extracts guards from the context literals of counterexamples-to-induction (CTIs), where PDR has already exposed a relational failure pattern. When individual CTIs expose only fragments of a larger control context, state-split search aggregates relational CTIs, selects a small proof-relevant state-region partition, and enumerates complete valuations of that partition as candidate guards. Both searches are heuristic candidate generators; acceptance remains the backend's proof-producing blocking procedure.

This paper makes three contributions:
\begin{itemize}
\item We identify guarded cross-copy equivalence as a useful proof shape for
hardware IFV, especially for designs where relational facts are tied to
control state rather than valid as unconditional invariants.

\item We present an in-engine guard-generation procedure that combines
CTI-local extraction with state-split search, producing guarded mismatch
obligations without user-supplied guarded lemmas.

\item We implement guarded search on ABC-PDR and rIC3. Across 12 IFV
benchmarks and two PDR backends, guarded predicates convert two contextual
baseline timeouts into completed proofs within 34.2--89.5s under an 1800s
limit, while reducing proof time by up to 10.8$\times$ on additional
benchmarks.
\end{itemize}

\begin{figure}[t]
\centering
\includegraphics[width=0.85\linewidth]{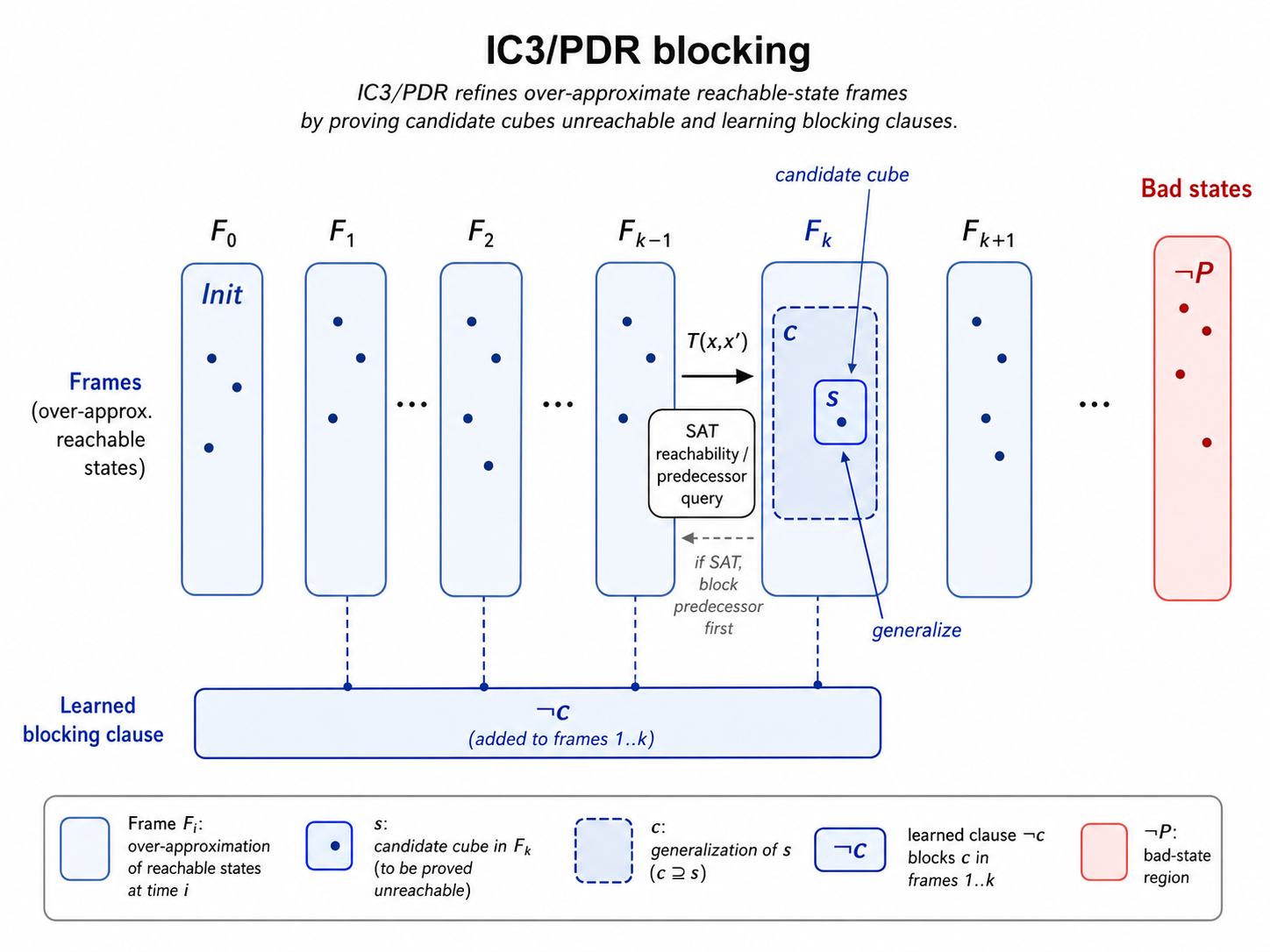}
\caption{IC3/PDR blocks an unreachable candidate cube $s$ by generalizing it
to a larger cube $c$ and adding the learned clause $\neg c$ to the frames.}
\label{fig:pdr_blocking}
\vspace{-5pt}
\end{figure}

\vspace{-3pt}
\section{Background}
\label{sec:background}
\vspace{-3pt}

IC3/PDR~\cite{Bradley2011IC3,Een2011abcpdr} is a SAT-based model checking
procedure for proving safety properties of finite-state transition systems. Let
the system be described by an initial-state formula $\mathit{Init}(x)$ and a
transition relation $T(x,x')$, where $x$ and $x'$ denote current- and next-state
variables. Given a safety property $P(x)$, IC3/PDR attempts to prove
\[
  \forall x.\; \mathit{Reach}(x) \Rightarrow P(x),
\]
or to return a counterexample trace reaching a bad state satisfying
$\neg P(x)$.

IC3/PDR maintains a sequence of clause sets, called frames,
\begin{align}
F_0, F_1, \ldots, F_k .
\end{align}
The first frame is the initial condition, $F_0=\mathit{Init}$, and later frames
over-approximate states reachable in increasing numbers of transitions. The
frames are maintained to satisfy monotonicity and relative inductiveness:
\begin{align}
F_i(x) \Rightarrow F_{i+1}(x),
\qquad
F_i(x) \wedge T(x,x') \Rightarrow F_{i+1}(x') .
\end{align}
When two adjacent frames become equivalent, the later frame is an inductive
invariant; if it excludes all bad states, the property is proved.

The main refinement step is cube blocking. A cube is a conjunction of state
literals. Given a candidate cube $s$ at frame $F_k$, IC3/PDR checks whether $s$
has a one-step predecessor in the previous frame:
\begin{align}
F_{k-1}(x) \wedge T(x,x') \wedge s(x') .
\label{eq:pdr_predecessor_query}
\end{align}
If this query is unsatisfiable, no state currently allowed by $F_{k-1}$ can
transition into $s$. IC3/PDR then generalizes $s$ to a larger unreachable cube
$c$ and learns the blocking clause $\neg c$, refining the relevant frames as
\begin{align}
F_i \leftarrow F_i \wedge \neg c
\qquad \text{for } 1 \leq i \leq k .
\end{align}
Fig.~\ref{fig:pdr_blocking} illustrates this process.
If the query in Eq.~\eqref{eq:pdr_predecessor_query} is satisfiable, the SAT solver returns a predecessor cube. IC3/PDR recursively attempts to block this predecessor at an earlier frame; if the recursion reaches $F_0$, the predecessor chain forms a real counterexample. Otherwise, clauses learned while blocking earlier predecessors may remove the path and allow the original cube to be blocked later.
This cube-blocking interface is the mechanism our method builds on. Rather
than changing the transition relation or the target safety property, our method generates additional proof obligations and submits them to the same standard PDR blocking procedure.

\begin{figure}[t]
\centering
\includegraphics[width=0.85\linewidth]{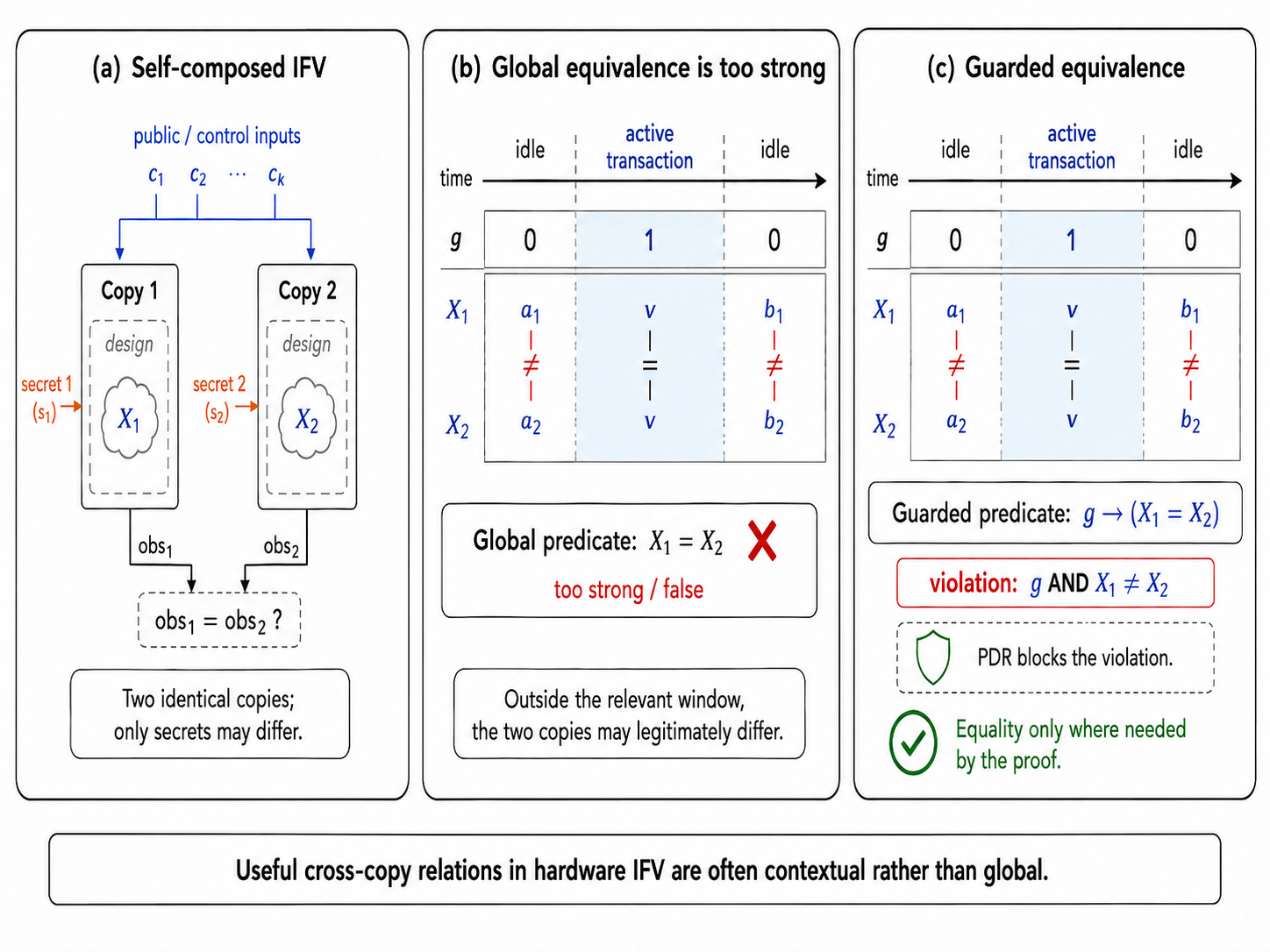}
\caption{Global equivalence can be too strong in hardware IFV. A guarded
equivalence predicate captures equality only in the control context where it
is required by the proof.}
\label{fig:motivation}
\vspace{-5pt}
\end{figure}


\vspace{-3pt}
\section{Motivation}
\label{sec}
\vspace{-3pt}

Self-composition turns IFV into a relational safety problem over two copies of
the same design, as shown in Fig.~\ref{fig:motivation}. This
structure naturally exposes cross-copy relationships~\cite{tan2026secic3}: if
an internal signal is determined only by public inputs and public state, the
corresponding signals in the two copies should agree. Making such relations
explicit can help IC3/PDR decompose the final output-equality proof into
smaller internal proof obligations.

However, many useful cross-copy equalities in hardware are contextual rather
than global. A signal may be meaningful only during a protocol phase,
transaction window, loop state, or valid cycle. Outside that context, it may
hold stale, inactive, or secret-dependent state that is irrelevant to the
observable behavior. Thus the unconditional claim that corresponding signals
are always equal can be too strong even when the IFV property is true.

Fig.~\ref{fig:motivation} illustrates this case. For corresponding
signals $X_1$ and $X_2$, equality may be required only when a control condition
$g$ holds. When $g$ is false, the protocol does not require $X$ to affect the
observable behavior, and the two copies may differ. The useful proof fact is
therefore not global equality of $X_1$ and $X_2$, but equality guarded by the
relevant control context.


This observation motivates guarded equivalence predicates. Instead of asking
IC3/PDR to use only equalities that hold everywhere, we pair a candidate
equality with a control or protocol condition. Operationally, the verifier
blocks the violating region in which the guard holds and the two copies
disagree. If the region is unreachable, the learned clauses become part of the
ordinary proof; if it is reachable, the candidate is discarded.

The remaining challenge is to discover the control contexts in which such guarded relations should be tested. A purely local strategy can reuse the context literals that already appear in PDR counterexamples-to-induction (CTIs), since these cubes expose where a relational proof attempt is currently failing. However, this still ties guard generation to a single local obligation. In FSM-rich designs, the relevant context may be distributed across multiple CTIs: PDR generalization may retain only partial state encodings, and a partial guard may combine control states where a mismatch is reachable with states where it should be ruled out. Our state-split search makes contextual discovery explicit. It aggregates relational CTIs, selects a small proof-relevant control partition, and enumerates complete valuations of that partition as candidate guarded regions. Thus, instead of waiting for PDR to encounter the right contextual relation inside an existing proof obligation, the verifier actively proposes guarded mismatch regions and asks the standard blocking procedure to prove or reject them.

\begin{figure}[t]
    \centering
    \includegraphics[width=0.8\linewidth]{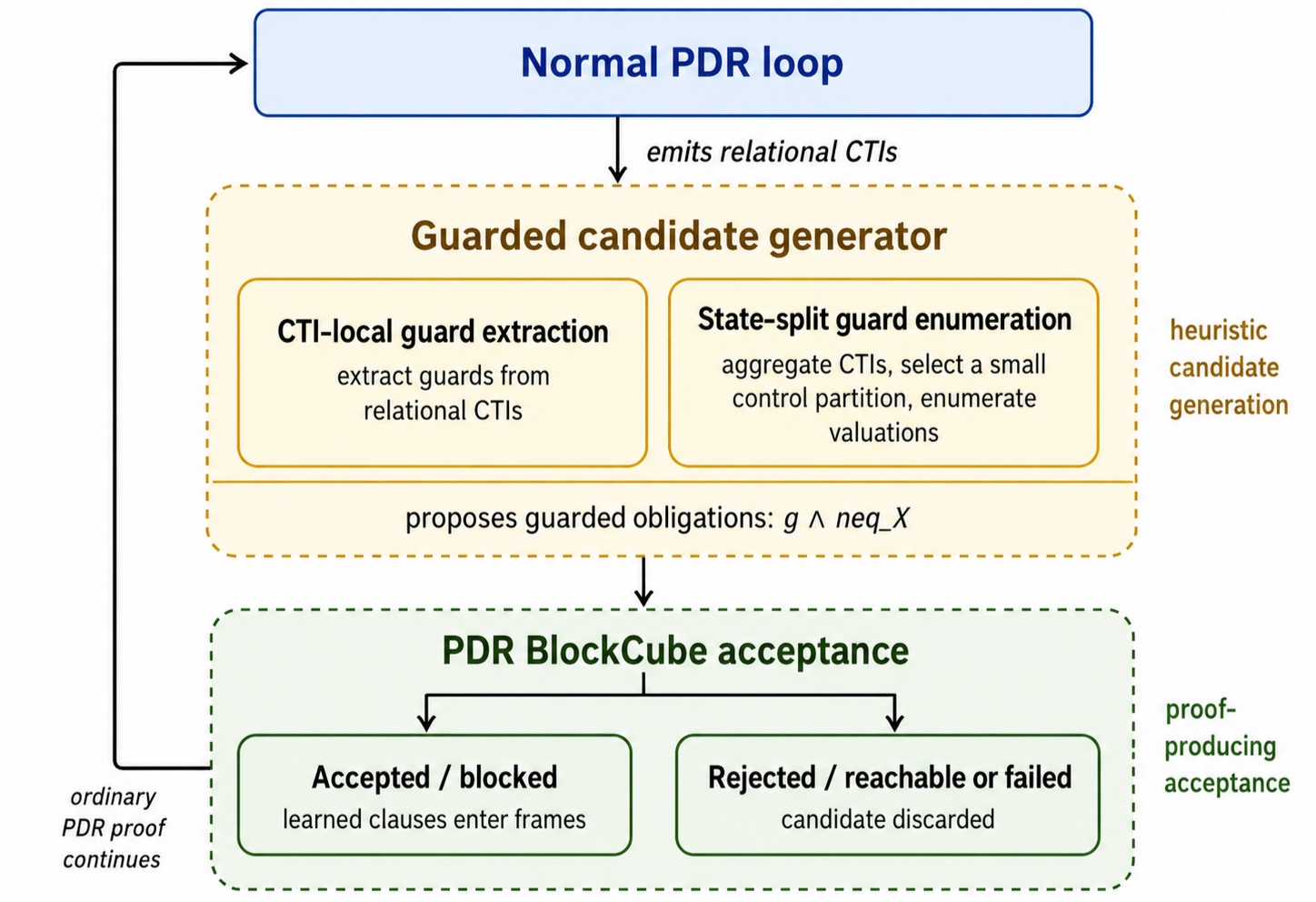}
    \caption{Overview of guarded search. Candidate generation is heuristic:
    CTI-local and state-split searches propose guarded mismatch obligations.
    Candidate acceptance remains proof-producing: blocked candidates contribute
    learned clauses to PDR frames, while failed or reachable candidates are
    discarded.}
    \label{fig:method_overview}
    \vspace{-5pt}
\end{figure}

\begin{table}[t]
\centering
\scriptsize
\setlength{\tabcolsep}{1.5pt}
\caption{Canonical guarded candidate-generation modes.}
\label{tab:method_guarded_modes}
\begin{tabularx}{\columnwidth}{@{}p{0.21\columnwidth}p{0.39\columnwidth}X@{}}
\toprule
Mode & Candidate family & Intended role \\
\midrule
$g$-AoN & $C(G_q,N_q)$ & cheap CTI-local bundle \\
$g$-Maximal & $C(G_q,\{\mathit{neq}_X\})$ & CTI-local singleton equality \\
$g$-Maximum & $C(G_q,S),\ S \subseteq N_q$ & bounded CTI-local subsets \\
$g$-StateSplit & $C(g_{K,a},\{\mathit{neq}_X\})$ & per-region singleton equality \\
$g$-Hybrid & schedule over modes & engineering policy \\
\bottomrule
\end{tabularx}
\end{table}

\begin{algorithm}[t]
\algtext*{EndIf}
\algtext*{EndFor}
\small
\caption{In-engine guarded candidate search}
\label{alg:method_guarded_search}
\begin{algorithmic}[1]
\Require Relational CTIs $\mathcal{Q}$, frame $k$, mode $M$, schedule $\mathcal{K}$
\State $\mathcal{C} \gets \emptyset$
\If{$M$ enables CTI-local search}
    \ForAll{$q \in \mathcal{Q}$ with positive mismatch predicates}
        \State $G_q \gets \textsc{ContextLits}(q)$
        \State $N_q \gets \textsc{MismatchPreds}(q)$
        \State $\mathcal{C} \gets \mathcal{C} \cup \textsc{LocalCands}(G_q,N_q,M)$
    \EndFor
\EndIf
\If{$M$ enables state-split search}
    \State rank non-predicate context variables using Eq.~\eqref{eq:method_freq}
    \ForAll{$K \in \mathcal{K}$}
        \State $S_K \gets \textsc{TopKControls}(K)$
        \ForAll{valuation $a$ of $S_K$}
            \ForAll{eligible mismatch predicate $\mathit{neq}_X$}
                \State $\mathcal{C} \gets \mathcal{C} \cup \{(S_K=a) \wedge \mathit{neq}_X\}$
            \EndFor
        \EndFor
    \EndFor
\EndIf
\ForAll{$c \in \mathcal{C}$}
    \If{$c$ intersects $\mathit{Init}$}
        \State discard $c$
    \ElsIf{\textsc{BlockCube}$(c,k)$ succeeds}
        \State keep the clauses learned by PDR
    \Else
        \State drop $c$
    \EndIf
\EndFor
\end{algorithmic}
\end{algorithm}

\section{Method: Guarded Equivalence Predicates}
\label{sec:method}



This section describes how guarded search generates and validates contextual cross-copy obligations inside PDR. The method observes relational CTIs produced by the normal PDR loop, proposes guarded mismatch cubes from those CTIs, and submits each candidate to the standard \textsc{BlockCube} procedure. Candidate generation is heuristic; candidate acceptance remains proof-producing.

Fig.~\ref{fig:method_overview} shows the flow. CTI-local search extracts guards from individual relational CTIs, while state-split search aggregates CTIs to construct candidate control regions. A candidate contributes to the proof only if \textsc{BlockCube} proves its guarded mismatch cube unreachable; otherwise it is discarded.


\subsection{Guarded Mismatch Obligations}

Consider corresponding internal signals ($X_1$) and ($X_2$) in the two copies of a self-composed design. We write
\begin{align}
\mathit{eq}_X \equiv (X_1 = X_2),
\qquad
\mathit{neq}_X \equiv (X_1 \neq X_2).
\end{align}
A guarded equality states that the two signals agree under a control context (g): $g \Rightarrow \mathit{eq}_X$ .
Rather than asserting this implication, we submit its violation to PDR:
\begin{align}
V(g,X) = g \wedge \mathit{neq}_X .
\end{align}
If \(V(g,X)\)  is blocked, the equality holds under (g) on all reachable states. If the cube is reachable or cannot be blocked, the candidate is discarded.
For a set of mismatch predicates ($\mathcal{N}$), we write
\begin{align}
C(g,\mathcal{N}) =
g \wedge \bigwedge_{\mathit{neq}_X \in \mathcal{N}} \mathit{neq}_X .
\end{align}
A singleton candidate directly proves a guarded equality when blocked. A bundled candidate only proves that a particular combination of mismatches cannot occur under the guard; it may still help PDR, but it does not by itself imply each individual equality.

\subsection{Relational CTI Capture}
\label{subsec:cti_capture}

Guarded search is driven by information already produced by the PDR engine. As PDR blocks bad states and their predecessors, it creates minimized cubes that explain proof failures and learned clauses. We observe these cubes and keep a bounded set of relational CTIs: cubes that contain at least one positive mismatch predicate $\mathit{neq}_X$. 
It stores evidence for later candidate generation but does not re-enter PDR, add a clause, or change the stock proof state.
A captured relational CTI $q$ is split into ordinary context literals and
mismatch predicates:
\begin{equation}
    q = G_q \cup N_q,
    \qquad
    N_q \subseteq \{\mathit{neq}_X\}_X .
    \label{eq:method_cti_split}
\end{equation}
The context ($G_q$) may include control-state bits, valid bits, counters, phase conditions, or other state literals. This capture is observational: it records evidence for candidate generation, but does not add clauses or modify the PDR state.

\subsection{Guarded Candidate Generation}
\label{subsec:candidate_generation}

The implementation uses two candidate sources. CTI-local search uses the state
region already exposed by one CTI. State-split search constructs a bounded
state-region partition from many relational CTIs. Both sources generate cubes;
neither source is trusted semantically.

\paragraph{CTI-local search}
For a relational CTI $q$, the context $G_q$ is used as a guard and the mismatch set $N_q$ supplies the candidate mismatches. The CTI-local family differs only in how it selects mismatch predicates under this guard; these selection policies follow the AoN, Maximal, and Maximum predicate-replacement modes of SecIC3~\cite{tan2026secic3}. The $g$-AoN mode submits one bundled candidate $C(G_q,N_q)$. The $g$-Maximal mode submits singleton candidates $C(G_q,\{\mathit{neq}_X\})$ for mismatches exposed by the CTI. The $g$-Maximum mode explores bounded subsets of $N_q$. 

\paragraph{State-split search}
A single CTI may expose only fragments of the useful state region. State-split
search therefore aggregates many relational CTIs and builds a small partition
of the proof state space. First, it excludes mismatch predicate latches from
consideration so that guards are built from ordinary state/context variables.
Then it ranks each remaining variable by how often it appears in CTI contexts:
\begin{equation}
    \mathit{freq}(x) =
    \left|\{q \in \mathcal{Q} : x \in \mathit{Vars}(G_q)\}\right| .
    \label{eq:method_freq}
\end{equation}
Both polarities of a variable contribute to the same frequency. For a parameter
$K$, the search selects the $K$ highest-ranked variables,
\begin{equation}
    S_K = \operatorname{TopK}_{x \in \mathcal{V}_{\mathit{ctx}}}
    \mathit{freq}(x),
    \label{eq:method_topk}
\end{equation}
and enumerates complete valuations of the selected partition. For a valuation
$a \in \{0,1\}^{|S_K|}$, the state-region guard is
\begin{equation}
    g_{K,a} = (S_K=a) =
    \bigwedge_{x_i \in S_K} (x_i=a_i).
    \label{eq:method_statesplit_guard}
\end{equation}
State-split search uses singleton mismatch obligations by default:
\begin{equation}
    (S_K=a) \wedge \mathit{neq}_X .
    \label{eq:method_statesplit_candidate}
\end{equation}
The selected variables are not trusted as a semantic FSM encoding; they only partition the state space into regions that PDR can test. Small (K) may produce coarse guards, while large (K) increases candidate count.

\begin{table*}[!t]
\centering
\caption{Runtime summary on the ABC-PDR backend with a 1800s budget.}
\label{tab:runtime_summary_abcpdr}
\vspace{-3pt}
{\footnotesize
\setlength{\tabcolsep}{5.5pt}
\setlength{\aboverulesep}{0.18ex}
\setlength{\belowrulesep}{0.18ex}
\setlength{\cmidrulekern}{0.25em}
\renewcommand{\arraystretch}{0.9}
\begin{tabular}{@{}lcc ccccccc@{}}
\toprule
& Base & SecIC3 & \multicolumn{7}{c}{Ours on ABC-PDR} \\
\cmidrule(lr){2-2}\cmidrule(lr){3-3}\cmidrule(lr){4-10}
Benchmark & PDR & Best & $g$-AoN & $g$-Maxl & $g$-Maxm & $g$-StSpl & $g$-Hyb & S+$g$-Maxl & S+$g$-StSpl \\
\midrule
\texttt{GCD}        & 10.7 & 7.2 & 1.8 & 1.7 & \textbf{1.3} & 5.8 & 1.7 & 3.3 & 37.7 \\
\texttt{FP\_ADD}    & \textsc{TO}:13 & 89.7 & 43.0 & 42.8 & 42.1 & 31.7 & 71.1 & 29.8 & \textbf{21.1} \\
\texttt{FP\_DIV}    & \textsc{TO}:59 & \textsc{TO}:65 & \textsc{TO}:66 & \textsc{TO}:66 & \textsc{TO}:66 & \textbf{80.0} & 136.8 & \textsc{TO}:59 & 122.1 \\
\texttt{normacc}    & \textsc{TO}:44 & \textsc{TO}:46 & \textsc{TO}:16 & \textsc{TO}:16 & \textsc{TO}:16 & \textbf{89.5} & \textsc{TO}:16 & \textsc{TO}:33 & \textsc{TO}:25 \\
\texttt{vctr}       & \textsc{TO}:63 & 444.6 & 64.7 & 42.5 & \textbf{41.6} & \textsc{TO}:27 & \textsc{TO}:43 & 64.9 & \textsc{TO}:40 \\
\texttt{Cache}      & 1745.3 & 150.8 & 174.8 & 169.9 & 170.1 & \textsc{TO}:16 & \textsc{TO}:17 & \textbf{63.6} & \textsc{TO}:16 \\
\texttt{Multiplier} & 6.1 & \textbf{0.1} & \textbf{0.1} & \textbf{0.1} & \textbf{0.1} & \textbf{0.1} & \textbf{0.1} & \textbf{0.1} & \textbf{0.1} \\
\texttt{Modexp}     & 0.6 & \textbf{0.4} & 0.6 & 0.6 & 0.6 & 0.6 & 0.9 & \textbf{0.4} & \textbf{0.4} \\
\texttt{Sodor}      & 1.4 & \textbf{0.4} & 0.8 & 0.9 & 0.7 & \textsc{TO}:6 & 0.7 & 0.5 & \textbf{0.4} \\
\texttt{SecEnclave} & 2.4 & \textbf{2.1} & 2.2 & 2.2 & 2.4 & 2.7 & 2.2 & \textbf{2.1} & 2.5 \\
\texttt{Rocket}     & 6.5 & \textbf{1.5} & 2.7 & 2.7 & 2.7 & 2.6 & 2.7 & 1.8 & 1.8 \\
\texttt{FP\_MUL}    & \textsc{TO}:13 & \textsc{TO}:15 & \textsc{TO}:12 & \textsc{TO}:12 & \textsc{TO}:12 & \textsc{TO}:9 & \textsc{TO}:12 & \textsc{TO}:11 & \textsc{TO}:9 \\
\bottomrule
\end{tabular}
}
\vspace{-8pt}
\end{table*}

\begin{table*}[!t]
\centering
\caption{Runtime summary on the rIC3 backend with a 1800s budget.}
\label{tab:runtime_summary_ric3}
\vspace{-3pt}
{\footnotesize
\setlength{\tabcolsep}{5.5pt}
\setlength{\aboverulesep}{0.18ex}
\setlength{\belowrulesep}{0.18ex}
\setlength{\cmidrulekern}{0.25em}
\renewcommand{\arraystretch}{0.9}
\begin{tabular}{@{}lcc ccccccc@{}}
\toprule
& Base & SecIC3 & \multicolumn{7}{c}{Ours on rIC3} \\
\cmidrule(lr){2-2}\cmidrule(lr){3-3}\cmidrule(lr){4-10}
Benchmark & rIC3 & Best & $g$-AoN & $g$-Maxl & $g$-Maxm & $g$-StSpl & $g$-Hyb & S+$g$-Maxl & S+$g$-StSpl \\
\midrule
\texttt{GCD}        & 8.0 & 1.0 & 1.8 & 0.8 & 1.4 & 0.7 & 0.9 & \textbf{0.5} & 0.6 \\
\texttt{FP\_ADD}    & 201.9 & 43.1 & 30.8 & 33.6 & 28.1 & 88.3 & 16.6 & \textbf{15.4} & 20.6 \\
\texttt{FP\_DIV}    & \textsc{TO}:54 & \textsc{TO}:99 & 38.1 & 37.2 & 37.6 & 51.0 & 209.7 & 37.1 & \textbf{34.2} \\
\texttt{normacc}    & \textsc{TO}:47 & \textsc{TO}:48 & \textsc{TO}:49 & \textsc{TO}:48 & \textsc{TO}:49 & \textsc{TO}:49 & \textbf{61.1} & \textsc{TO}:49 & \textsc{TO}:49 \\
\texttt{vctr}       & 250.3 & 82.8 & \textbf{7.7} & 10.8 & 12.1 & 42.2 & 16.5 & 9.5 & 28.0 \\
\texttt{Cache}      & \textsc{TO}:15 & 69.4 & 8.8 & 7.3 & 12.3 & 9.2 & 96.7 & \textbf{6.8} & \textbf{6.8} \\
\texttt{Multiplier} & 2.1 & \textbf{0.1} & \textbf{0.1} & \textbf{0.1} & \textbf{0.1} & 2.2 & \textbf{0.1} & \textbf{0.1} & 1.1 \\
\texttt{Modexp}     & 0.3 & \textbf{0.2} & 0.6 & 0.6 & 0.6 & 0.4 & 0.6 & 0.3 & 0.3 \\
\texttt{Sodor}      & 2.0 & \textbf{1.1} & 4.0 & 3.5 & 4.1 & 3.4 & 3.7 & 3.4 & 3.0 \\
\texttt{SecEnclave} & 28.0 & \textbf{27.5} & 28.2 & 29.8 & 41.2 & 29.0 & 29.0 & 33.5 & 33.1 \\
\texttt{Rocket}     & 8.3 & \textbf{5.4} & 164.2 & 123.2 & 226.5 & 182.0 & 189.7 & \textsc{TO}:620 & 147.3 \\
\texttt{FP\_MUL}    & \textsc{TO}:14 & \textsc{TO}:15 & \textsc{TO}:10 & \textsc{TO}:10 & \textsc{TO}:10 & \textsc{TO}:10 & \textsc{TO}:10 & \textsc{TO}:10 & \textsc{TO}:11 \\
\bottomrule
\end{tabular}
}
\vspace{-8pt}
\end{table*}

Table~\ref{tab:method_guarded_modes} summarizes the canonical modes used in
our evaluation. Hybrid scheduling is included for completeness: it runs a cheap CTI-local phase first and can escalate to state-split search, but it is an engineering policy rather than a separate proof rule.
Algorithm~\ref{alg:method_guarded_search} gives the in-engine search. The
algorithm intentionally separates proposal from acceptance: all candidates are
ordinary cubes, and only \textsc{BlockCube} can turn a candidate into learned
PDR clauses.

\vspace{-1pt}
\subsection{Candidate Acceptance and Soundness}
\label{subsec:method_soundness}

Guarded candidates are auxiliary proof obligations, not assumptions. A candidate contributes to the proof only if the backend proves its guarded mismatch cube unreachable using the ordinary PDR blocking procedure and learns blocking clauses. Candidates that intersect initial states, are reachable, or fail to block are discarded. Therefore, guard generation can affect which auxiliary cubes are attempted, but not the transition relation, the initial condition, the IFV property, or the proof rule used to accept learned clauses. CTI-local search, state-split search, and hybrid scheduling are untrusted proposal mechanisms; soundness rests on the backend's blocking and final proof/certification path.

\vspace{-3pt}
\section{Evaluation}
\label{sec:evaluation}
\vspace{-3pt}

\subsection{Experimental Setup}
\label{subsec:eval_setup}
\vspace{-1pt}

We implemented guarded candidate generation in two PDR-based IFV flows: an ABC-PDR backend and an rIC3 backend. Both backends support the same guarded families: \(g\)-AoN, \(g\)-Maximal, \(g\)-Maximum, \(g\)-StateSplit, and \(g\)-Hybrid, together with selected symmetry-enabled combinations. 
All experiments were run on a machine with an Intel Core i9-14900K CPU and 188 GiB of RAM. Each proof process is single-threaded. Independent benchmark/mode runs were executed as parallel processes for throughput, and reported runtimes are per-process wall-clock times. Each run uses a 1800s timeout.

\vspace{-1pt}
\subsection{Benchmarks and Baselines}
\label{subsec:eval_benchmarks}
\vspace{-1pt}



We evaluate 12 IFV benchmarks: ten from the SecIC3 benchmark suite~\cite{tan2026secic3} and two diagnostic designs, \texttt{vctr} and \texttt{normacc}, that stress contextual guarded relations. The SecIC3 benchmarks cover arithmetic datapaths, cryptographic/control-heavy designs, and processor fragments. The diagnostic designs isolate cases where the relevant cross-copy relation is tied to a control context rather than useful as an unconditional invariant.
We compare against the unmodified backend and the SecIC3-style predicate-replacement modes AoN, Maximal, and Maximum, with and without copy-swap symmetry. The SecIC3-best column reports the best result among the seven reproduced SecIC3-style configurations on the same backend. Guarded modes use the same backends, but add auxiliary guarded mismatch obligations submitted to the standard PDR blocking procedure.

\vspace{-1pt}
\subsection{Runtime Results}
\vspace{-1pt}

Tables~\ref{tab:runtime_summary_abcpdr} and~\ref{tab:runtime_summary_ric3}
report runtime summaries for ABC-PDR~\cite{Brayton2010abc} and rIC3~\cite{Su2025rIC3}. Each
row is a benchmark. 
Entries are wall-clock seconds for completed proofs; \textsc{TO}:$n$ denotes a timeout after reaching frame $n$. Bold entries mark the fastest completed proof in each row.

Guarded search closes contextual cases that the best reproduced SecIC3-style baseline does not prove under the same 1800s budget. On ABC-PDR, guarded modes prove \texttt{FP\_DIV} and \texttt{normacc} in 80.0s and 89.5s, respectively, where SecIC3-best times out. On rIC3, the corresponding proofs complete in 34.2s and 61.1s. Thus, guarded obligations do not merely improve runtime on easy cases; they turn several baseline timeouts into completed proofs.

Guarded search also accelerates already-provable benchmarks. Across the two backends, the best guarded configuration improves \texttt{GCD}, \texttt{FP\_ADD}, \texttt{vctr}, and \texttt{Cache} by up to 10.8$\times$ over SecIC3-best. The benefit is not uniform: easy or non-contextual cases leave little room for improvement, and guarded candidates can add overhead. \texttt{FP\_MUL} remains unsolved on both backends, suggesting a separate bottleneck in wide arithmetic
or operator reasoning rather than in contextual guard discovery.


\vspace{-1pt}
\subsection{Mode Analysis}
\label{subsec:eval_modes}
\vspace{-1pt}

No guarded mode is uniformly best across benchmarks or backends. CTI-local modes work well when a single relational CTI already exposes the relevant control context; this explains their gains on cases such as \texttt{GCD}, \texttt{vctr}, and \texttt{Cache}. State-split search is useful when the needed context is distributed across multiple CTIs. By selecting frequently appearing context variables and enumerating complete valuations, it avoids overly coarse partial guards and is essential for the ABC-PDR proofs of \texttt{FP\_DIV} and \texttt{normacc}. Symmetry and guarded search are complementary but not universally beneficial: their combination helps on some cases, including \texttt{FP\_ADD} and \texttt{Cache}, but no fixed combination dominates. Similarly, g-Hybrid is only a scheduling policy; it can help when it reaches the right guarded family early, but it can also add overhead. We therefore report individual guarded modes rather than presenting a single default configuration.

\section{Conclusion}
\label{sec:conclusion}


This paper introduced guarded equivalence predicates for PDR-based hardware information-flow verification. The method tests whether a cross-copy mismatch is unreachable under a candidate control context, and accepts guards only through the backend's standard blocking procedure. Guards are proposed from relational CTIs using CTI-local extraction and state-split search. Experiments on ABC-PDR and rIC3 show that guarded obligations close contextual IFV cases missed by baseline equivalence predicates and accelerate additional benchmarks, making them a practical complement to symmetry and equivalence-based reasoning.

\section*{Acknowledgment}
This work was partially supported by a grant from the CHEST IUCRC.

\bibliographystyle{IEEEtran}
\bibliography{refs}

\begin{thebibliography}{10}
\providecommand{\url}[1]{#1}
\csname url@samestyle\endcsname
\providecommand{\newblock}{\relax}
\providecommand{\bibinfo}[2]{#2}
\providecommand{\BIBentrySTDinterwordspacing}{\spaceskip=0pt\relax}
\providecommand{\BIBentryALTinterwordstretchfactor}{4}
\providecommand{\BIBentryALTinterwordspacing}{\spaceskip=\fontdimen2\font plus
\BIBentryALTinterwordstretchfactor\fontdimen3\font minus \fontdimen4\font\relax}
\providecommand{\BIBforeignlanguage}[2]{{%
\expandafter\ifx\csname l@#1\endcsname\relax
\typeout{** WARNING: IEEEtran.bst: No hyphenation pattern has been}%
\typeout{** loaded for the language `#1'. Using the pattern for}%
\typeout{** the default language instead.}%
\else
\language=\csname l@#1\endcsname
\fi
#2}}
\providecommand{\BIBdecl}{\relax}
\BIBdecl

\bibitem{Gleissenthall2021constranhwverification}
\BIBentryALTinterwordspacing
K.~v.~Gleissenthall, R.~G. K\i{}c\i{}, D.~Stefan, and R.~Jhala, ``Solver-aided constant-time hardware verification,'' in \emph{Proceedings of the 2021 ACM SIGSAC Conference on Computer and Communications Security}, ser. CCS '21.\hskip 1em plus 0.5em minus 0.4em\relax New York, NY, USA: Association for Computing Machinery, 2021, p. 429–444. [Online]. Available: \url{https://doi.org/10.1145/3460120.3484810}
\BIBentrySTDinterwordspacing

\bibitem{Lou2022hw_security_attacks}
\BIBentryALTinterwordspacing
X.~Lou, T.~Zhang, J.~Jiang, and Y.~Zhang, ``A survey of microarchitectural side-channel vulnerabilities, attacks, and defenses in cryptography,'' \emph{ACM Comput. Surv.}, vol.~54, no.~6, Jul. 2021. [Online]. Available: \url{https://doi.org/10.1145/3456629}
\BIBentrySTDinterwordspacing

\bibitem{Kocher2019SepctreMeltdown}
P.~Kocher, J.~Horn, A.~Fogh, D.~Genkin, D.~Gruss, W.~Haas, M.~Hamburg, M.~Lipp, S.~Mangard, T.~Prescher, M.~Schwarz, and Y.~Yarom, ``Spectre attacks: Exploiting speculative execution,'' in \emph{2019 IEEE Symposium on Security and Privacy (SP)}, 2019, pp. 1--19.

\bibitem{Hu2022HIFT}
\BIBentryALTinterwordspacing
W.~Hu, A.~Ardeshiricham, and R.~Kastner, ``Hardware information flow tracking,'' \emph{ACM Comput. Surv.}, vol.~54, no.~4, May 2021. [Online]. Available: \url{https://doi.org/10.1145/3447867}
\BIBentrySTDinterwordspacing

\bibitem{Ardeshiricham2017RTLIFT}
A.~Ardeshiricham, W.~Hu, J.~Marxen, and R.~Kastner, ``Register transfer level information flow tracking for provably secure hardware design,'' in \emph{Design, Automation \& Test in Europe Conference \& Exhibition (DATE), 2017}, 2017, pp. 1691--1696.

\bibitem{Hu2011GLIFT}
W.~Hu, J.~Oberg, A.~Irturk, M.~Tiwari, T.~Sherwood, D.~Mu, and R.~Kastner, ``Theoretical fundamentals of gate level information flow tracking,'' \emph{IEEE Transactions on Computer-Aided Design of Integrated Circuits and Systems}, vol.~30, no.~8, pp. 1128--1140, 2011.

\bibitem{Gleissenthall2019IODINE}
K.~V. Gleissenthall, R.~G. Kici, D.~Stefan, and R.~Jhala, ``Iodine: verifying constant-time execution of hardware,'' in \emph{Proceedings of the 28th USENIX Conference on Security Symposium}, ser. SEC'19.\hskip 1em plus 0.5em minus 0.4em\relax USA: USENIX Association, 2019, p. 1411–1428.

\bibitem{Barthe2004selfcomposition}
G.~Barthe, P.~D'Argenio, and T.~Rezk, ``Secure information flow by self-composition,'' in \emph{Proceedings. 17th IEEE Computer Security Foundations Workshop, 2004.}, 2004, pp. 100--114.

\bibitem{Clarkson2010Hyperproperty}
M.~R. Clarkson and F.~B. Schneider, ``Hyperproperties,'' \emph{J. Comput. Secur.}, vol.~18, no.~6, p. 1157–1210, Sep. 2010.

\bibitem{tan2026secic3}
\BIBentryALTinterwordspacing
Q.~Tan, A.~Gaonkar, Y.-W. Fan, A.~Gupta, and S.~Malik, ``Secic3: Customizing ic3 for hardware security verification,'' 2026. [Online]. Available: \url{https://arxiv.org/abs/2601.21353}
\BIBentrySTDinterwordspacing

\bibitem{Bradley2011IC3}
A.~R. Bradley, ``Sat-based model checking without unrolling,'' in \emph{Proceedings of the 12th International Conference on Verification, Model Checking, and Abstract Interpretation}, ser. VMCAI'11.\hskip 1em plus 0.5em minus 0.4em\relax Berlin, Heidelberg: Springer-Verlag, 2011, p. 70–87.

\bibitem{Een2011abcpdr}
N.~Een, A.~Mishchenko, and R.~Brayton, ``Efficient implementation of property directed reachability,'' in \emph{2011 Formal Methods in Computer-Aided Design (FMCAD)}, 2011, pp. 125--134.

\bibitem{Brayton2010abc}
\BIBentryALTinterwordspacing
R.~Brayton and A.~Mishchenko, ``Abc: an academic industrial-strength verification tool,'' in \emph{Proceedings of the 22nd International Conference on Computer Aided Verification}, ser. CAV'10.\hskip 1em plus 0.5em minus 0.4em\relax Berlin, Heidelberg: Springer-Verlag, 2010, p. 24–40. [Online]. Available: \url{https://doi.org/10.1007/978-3-642-14295-6_5}
\BIBentrySTDinterwordspacing

\bibitem{Su2025rIC3}
\BIBentryALTinterwordspacing
Y.~Su, Q.~Yang, Y.~Ci, T.~Bu, and Z.~Huang, ``The ric3 hardware model checker,'' in \emph{Computer Aided Verification: 37th International Conference, CAV 2025, Zagreb, Croatia, July 23–25, 2025, Proceedings, Part I}.\hskip 1em plus 0.5em minus 0.4em\relax Berlin, Heidelberg: Springer-Verlag, 2025, p. 185–199. [Online]. Available: \url{https://doi.org/10.1007/978-3-031-98668-0_9}
\BIBentrySTDinterwordspacing

\end{thebibliography}

\end{document}